# High-temperature quantum oscillations caused by recurring Bloch states in graphene superlattices

R. Krishna Kumar[1,2,3], X. Chen[2], G. H. Auton[2], A. Mishchenko[1], D. A. Bandurin[1], S. V. Morozov[4,5], Y. Cao[2], E. Khestanova[1], M. Ben Shalom[1], A. V. Kretinin[2], K. S. Novoselov[2], L. Eaves[2,6], I. V. Grigorieva[1], L. A. Ponomarenko[3], V. I. Fal'ko[1,2], A. K. Geim[1,2]

[1]School of Physics & Astronomy, University of Manchester, Oxford Road, Manchester M13 9PL, United Kingdom
[2]National Graphene Institute, University of Manchester, Manchester M13 9PL, United Kingdom
[3]Department of Physics, University of Lancaster, Lancaster LA1 4YW, United Kingdom
[4]Institute of Microelectronics Technology and High Purity Materials, RAS, Chernogolovka 142432, Russia
[5]National University of Science and Technology 'MISiS', Leninsky Pr. 4, 119049 Moscow, Russia
[6]School of Physics and Astronomy, University of Nottingham NG7 2RD, United Kingdom

**Cyclotron motion of charge carriers in metals and semiconductors leads to Landau quantization and magneto-oscillatory behavior in their properties. Cryogenic temperatures are usually required to observe these oscillations. We show that graphene superlattices support a different type of quantum oscillations that do not rely on Landau quantization. The oscillations are extremely robust and persist well above room temperature in magnetic fields of only a few T. We attribute this phenomenon to repetitive changes in the electronic structure of superlattices such that charge carriers experience effectively no magnetic field at simple fractions of the flux quantum per superlattice unit cell. Our work points at unexplored physics in Hofstadter butterfly systems at high temperatures.**

Magneto-oscillations are a well-known and important phenomenon in condensed matter physics. Despite having a variety of experimental manifestations, there are only a few basic types of oscillations, either of quantum or semiclassical origin[1-5]. Semiclassical size effects such as, e.g., Gantmakher and Weiss oscillations appear due to commensurability between the cyclotron orbit and a certain length in an experimental system[1-4]. Quantum magneto-oscillations are different in that they arise from periodic changes in the interference along closed electron trajectories[1-5]. Most commonly, quantum oscillations involve cyclotron trajectories. This leads to Landau quantization and, consequently, Shubnikov-de Haas (SdH) oscillations in magnetoresistance and the associated oscillatory behavior in many other properties[1-3]. In addition, quantum oscillations may arise from interference on trajectories imposed by sample geometry, leading to, e.g., Aharonov-Bohm oscillations in mesoscopic rings[3,5]. Whatever their exact origin, the observation of such oscillatory effects normally requires low temperatures ($T$), and this requirement is particularly severe in the case of quantum oscillations that rely on the monochromaticity of interfering electron waves. Even in graphene with its massless Dirac spectrum and exceptionally large cyclotron gaps, SdH oscillations rarely survive above 100 K. At room $T$, high magnetic fields $B$ of about 30 T are needed to observe the last two SdH oscillations arising from the maximal gaps between zeroth and first Landau levels (LLs) of graphene[6]. In all other materials, quantum oscillations disappear at much lower temperatures.

Electronic systems with superlattices can also exhibit magneto-oscillations. In this case, the interference of electrons diffracting at a superlattice potential in a magnetic field results in fractal, self-similar spectra that are often referred to as Hofstadter butterflies (HB) (refs. 7-12). Their fractal structure reflects the fact that charge carriers effectively experience no magnetic field if magnetic flux $\phi$ through the superlattice unit cell is commensurate with the magnetic flux quantum, $\phi_0$ (refs. 7-9). This topic has attracted intense interest for decades[11-16] but received a particular boost due to the recent observation of clear self-similar features in transport characteristics and in the density of states (DoS) of graphene superlattices[17-25]. Because the Hofstadter butterfly depicts quantum states developed from partial admixing of graphene's original LLs[12], superlattice-related gaps become smeared at relatively low $T$, well below those at which signatures of quantization in the main spectrum disappear. Therefore, it is perhaps not surprising that so far no investigation of Hofstadter systems was attempted beyond the low $T$ regime. As shown below, this has caused important physics to be overlooked: Superlattices exhibit robust high-$T$ oscillations in their transport characteristics, which are different in origin from the known oscillatory effects.

Our transport measurements were carried out using multiterminal Hall bar devices (fig. S1) made from graphene superlattices as described in Supplementary Information[26]. In brief, monolayer graphene was placed on top of a hexagonal boron nitride (hBN) crystal and their crystallographic axes were aligned with an accuracy of better than 2° (refs. 17-24). The resulting moiré pattern gives rise to a periodic potential that is known to affect the electronic spectrum of graphene[23-25]. To ensure that the charge carriers have high mobility, the graphene was encapsulated using a second hBN crystal, which was intentionally misaligned by ~15° with respect to graphene's axes. Although the second hBN layer also leads to a moiré pattern, it has a short periodicity and, accordingly, any superlattice effects may appear only at high carrier concentrations $n$ or ultra-high $B$, beyond those accessible in transport experiments[17-25]. Therefore, the second hBN effectively serves as an inert atomically-flat cover protecting graphene from the environment. Six superlattice devices were investigated and showed the behavior described below. As a reference, we also studied devices made following the same fabrication procedures but with the graphene misaligned with respect to both top and bottom hBN layers.

Fig. 1a shows typical behavior of the longitudinal resistivity $\rho_{xx}$ for graphene superlattices as a function of $B$ at various $T$. For comparison, Fig. 1b plots similar measurements for the reference device. In the latter case, $\rho_{xx}$ exhibits pronounced SdH oscillations at liquid-helium $T$, which develop into the quantum Hall effect above a few T. The SdH oscillations are rapidly suppressed with increasing temperature and completely vanish above liquid-nitrogen $T$, the standard behavior for graphene in these relatively weak fields[28,29]. In stark contrast, graphene superlattices exhibit prominent oscillations over the entire $T$ range (Fig. 1a and fig. S2). At both high and low $T$, the oscillations are periodic in $1/B$ (figs. S3-S4). One can see that the oscillations in Fig. 1a change their frequency at approximately 50 K. This is the same $T$ range where SdH oscillations disappear in the reference device of Fig. 1b. For certain ranges of $n$ and $B$, we observed that SdH oscillations seemed to vanish first, before new oscillations emerged at higher $T$. An example of such nonmonotonic temperature dependence is shown in fig. S2. To emphasize the robustness of the high-$T$ oscillations, we show that they remain well developed even at boiling-water $T$ in moderate $B$ (Fig. 1c). The oscillations were observed even at higher $T$ but, above 400 K, our devices (both superlattice and reference ones) showed rapid deterioration in quality and strong hysteresis as a function of gate voltage.

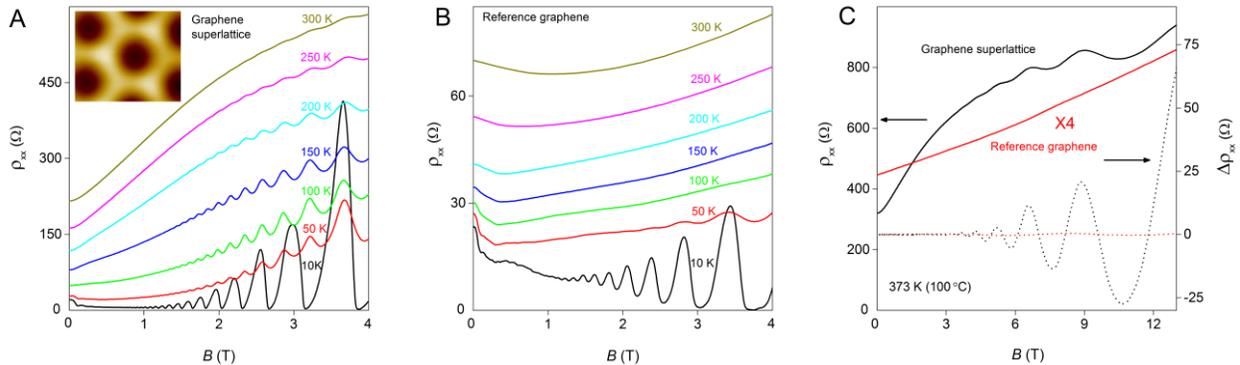

**Figure 1| High-*T* oscillations in graphene superlattices. a,** $\rho_{xx}$ at relatively small *B* for a superlattice device with the moiré periodicity of ≈13.6 nm. The electron density is $n \approx 1.7 \times 10^{12}$ cm$^{-2}$ and induced by a back gate voltage. Inset: Scanning tunneling image illustrates typical moiré patterns found in our devices [for details, see (ref. *27*)]. **b,** As in (a) but using the reference device at the same *n*. **c,** Magnetoresistance curves for devices in (a) and (b) at 100°C (solid curves). $\rho_{xx}$ for the reference device is multiplied by 4. Dotted curves: The oscillatory behavior is emphasized by subtracting the backgrounds described by 4th order polynomials. Graphene superlattices exhibit oscillations starting below 4 T whereas no sign of oscillatory behavior could be discerned at this *T* in our reference devices.

The high-*T* and SdH oscillations differ not only in their periodicity and thermal stability but also have distinctly different *n* dependences. Figs. 2a-b show Landau fan diagrams for the longitudinal conductivity $\sigma_{xx}$ of graphene superlattices as a function of *B* and *n* (we plot $\sigma_{xx}$ rather than $\rho_{xx}$ to facilitate the explanation given below for the origin of the high-*T* oscillations). At low *T* (Fig. 2a), we observe the same behavior as reported previously[17-22]: Numerous LLs fan out from the main (*n* = 0) and second-generation neutrality points (NPs) that are found at $n = \pm n_0$ where $n_0 = 4/S$ corresponds to 4 charge carriers per superlattice unit cell with the area $S = \sqrt{3}a^2/2$ and the superlattice period *a* (refs. 17-22). The LL intersections result in third-generation NPs at finite *B* (ref. 19). Minima in $\sigma_{xx}$ evolve linearly in *B* and originate from first-, second- and third-generation NPs (refs. 17-22). This reflects the fact that the DoS for all LLs (including those due to fractal gaps) is the same and proportional to *B* (ref. 11). At *T* > 100 K, the Landau quantization dominating the low-*T* diagrams wanes and, instead, there emerge oscillations with a periodicity independent of *n* (Figs. 2b-c). This independence on *n* clearly distinguishes the high-*T* oscillations from all the known magneto-oscillatory effects arising from either Landau quantization or commensurability[1-5]. For the reasons that become clear below, we refer to the found high-*T* phenomenon as Brown-Zak (BZ) oscillations[7,8].

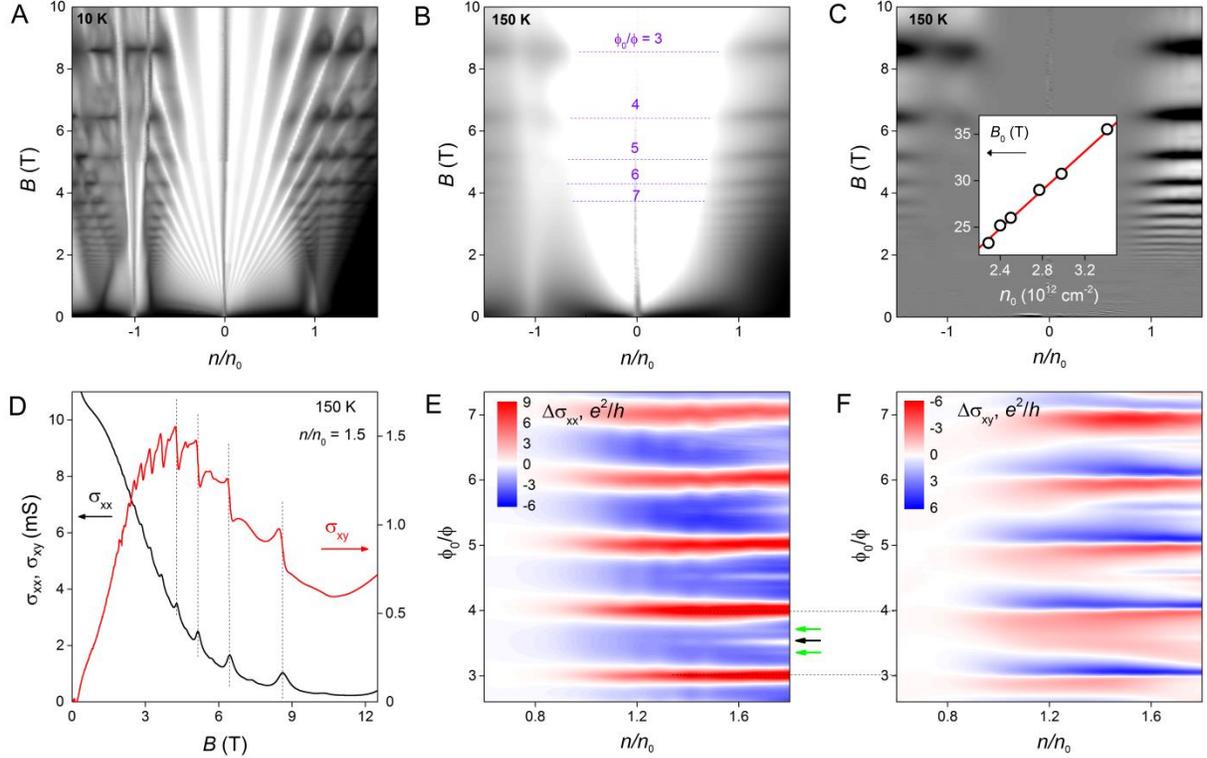

**Figure 2| Concentration and field dependence of Brown-Zak oscillations. a,** Low-$T$ fan diagram $\sigma_{xx}(n,B)$ for a superlattice device with $a \approx 13.9$ nm. The grey scale is logarithmic: white 0.015 mS; black 15 mS. **b,** Same device at 150 K. Logarithmic grey scale: white 0.1 mS; black 10 mS. The dotted lines mark $B = \phi_0/qS$. **c,** Same as (b) but for $\Delta\sigma_{xx}$ (that is, after subtracting a smooth background). Linear grey scale: $\pm 0.3$ mS. Inset: Fundamental frequency $B_0$ of BZ oscillations found in our devices as a function of $n_0 = 8/\sqrt{3}a^2$. **d,** Near $B = \phi_0/qS$ (dashed lines are for $q$ = 3 to 6), local changes in $\sigma_{xx}$ and $\sigma_{xy}$ resemble magnetotransport in metals near zero field. **e,** Part of (c) near the second-generation NP for electron doping is magnified and plotted as a function of $\phi_0/\phi$. Main maxima in $\Delta\sigma_{xx}$ occur at $\phi_0/\phi = q$. A few extra maxima for $p$ = 2 and 3 are indicated by black and green arrows, respectively (also, see fig. S6). **f,** Corresponding behavior of $\Delta\sigma_{xy}$ (smooth background subtracted). Its zeros align with the red maxima in $\Delta\sigma_{xx}$.

BZ oscillations become stronger with increasing doping (Fig. 2c), in agreement with the fact that the superlattice spectrum is stronger modified at energies away from the main Dirac point[17-22]. The maxima in $\sigma_{xx}$ are found at $B = \phi_0/qS$, which corresponds to unit fractions of $\phi_0$ piercing a superlattice unit cell ($q$ is integer). The relation between the superlattice period and the periodicity of the high-$T$ oscillations holds accurately for all our devices (inset of Fig. 2c and, also, fig. S3). This is the same periodicity that underlines the Hofstadter spectrum and describes the recurrence of third-generation NPs (refs. 17 & 18). However, BZ oscillations emerge most profoundly at high $T$, in the absence of any remaining signs of the Hofstadter spectrum or even Landau quantization (fig. S5). Importantly, our capacitance measurements (fig. S5a) reveal no sign of behavior similar to BZ oscillations in the DoS, even at liquid-helium $T$ that allow clearest view of the HB (refs. 20-25). These observations show that BZ oscillations are a transport phenomenon, unrelated to the spectral gaps that make up the Hofstadter spectrum. Note that BZ oscillations do not disappear at low $T$ and, retrospectively, can be recognized as horizontal streaks on the transport Landau fan diagrams[17-19], which connect third-generation NPs (Fig. 2a). However, the streaks are heavily crisscrossed by LLs, which makes them easy to overlook or wrongly associate with the quantized Hofstadter spectrum[17].

It is interesting to note that the HB spectrum is expected to exhibit a fractal periodicity associated with not only unit but all the simple fractions, $p/q$, corresponding to $p$ flux quanta per $q$ cells. No signatures of such higher-order states were found in the previous experiments[17-22], nor can be resolved in our present fan diagrams at low $T$. However, the fractions with $p$ = 2 and 3 become evident in BZ oscillations (Fig. 2e) and are most prominent at high $n$ (fig. S6). This again indicates the same underlying periodicity for BZ oscillations as the one governing the HB spectrum. We also find that BZ oscillations are stronger for electrons than for holes (Fig. 2b and fig. S7). This is in contrast to the relative strengths of all the other features reported previously for graphene-on-hBN superlattices[17-25]. The electron-hole asymmetry is probably connected to the observed stronger electron-phonon scattering for hole doping (fig. S7).

To explain BZ oscillations, we recall that, at $B = \phi_0 p/Sq$, the electronic spectrum of superlattices can be reduced to the case of zero magnetic field by introducing new Bloch states and the associated 'magnetic' minibands, different for each $p/q$. This observation was put forward by Brown[7] and Zak[8] and predates the work by Hofstadter[10]. Examples of BZ minibands for several unit fractions of $\phi_0$ are shown in Fig. 3 and fig. S8, using a generic graphene-on-hBN potential[26]. Each miniband can be viewed as a 'superlattice-broadened' LL such that its energy dispersion $\varepsilon(\vec{k})$ disappears in the limit of vanishing superlattice modulation[12]. If the Fermi energy $\varepsilon_F$ lies within these superlattice-broadened LLs, the system should exhibit a metallic behavior[25]. The Hofstadter spectrum can then be understood as Landau quantization of BZ minibands in the effective field $B_{eff} = B - \phi_0 p/qS$ (refs. 20 & 30). With this concept in mind, let us take a closer look at the experimental behavior of $\sigma_{xx}$ and the Hall conductivity $\sigma_{xy}$ at high $T$ and small $B_{eff}$, that is, in the absence of Landau quantization in BZ minibands (Figs. 2d-f and fig. S6). One can see that every time BZ minibands are formed, $\sigma_{xx}$ exhibits a local maximum and $\sigma_{xy}$ shows a smeared $B_{eff}/(1+B_{eff}^2)$-like feature on top of a smoothly varying background. This local behavior resembles changes in $\sigma_{xx}$ and $\sigma_{xy}$ expected for any metallic system near zero $B$ and approximated by the functional forms $1/(1+B^2)$ and $B/(1+B^2)$, respectively[1-3]. This behavior matches rather accurately the shape of local changes in $\sigma_{xx}$ and $\sigma_{xy}$ near fractional fluxes $\phi = \phi_0 p/q$, which correspond to $B_{eff}$ =0 (Figs. 2d & 3; fig. S6).

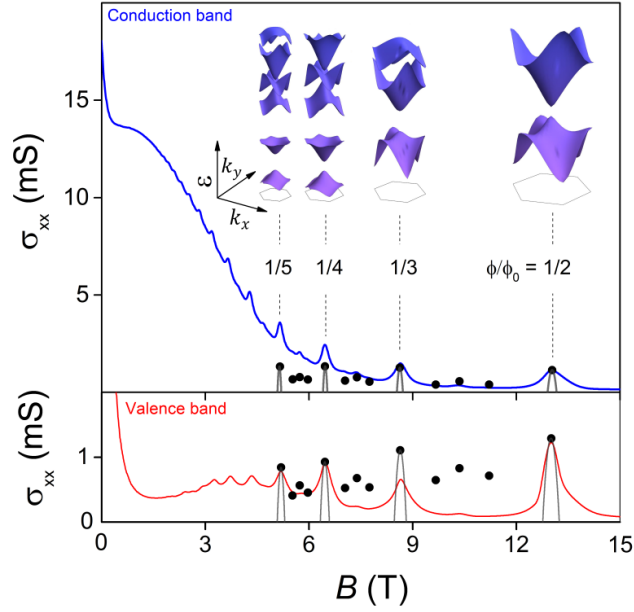

**Figure 3| BZ oscillations as recurring Bloch states in small effective fields.** Solid curves: $\sigma_{xx}$ at 100 K for electron and hole doping ($n/n_0 = \pm 1.6$; top and bottom panels, respectively) in a superlattice device with $a \approx 13.6$ nm. Black dots: $\sigma_{xx}$ calculated in the constant-$\tau$ approximation for different $p$ and $q$. Thin grey parabolas are calculated $\sigma_{xx}(B_{eff})$. Inset: BZ minibands $\varepsilon(\vec{k})$ inside the first Brillouin zones indicated by the grey hexagons (their size decreases with increasing $q$). The minibands were calculated for a generic graphene-on-hBN superlattice[30] and, for example, correspond to broadened LLs from 2 to 3 for $q = 2$ and from 3 to 8 for $q = 5$. Shown are only minibands at energies relevant to the doping level on the experimental curves.

The described analogy between magnetotransport in normal metals and in BZ minibands can be elaborated using the approximation of a constant scattering time $\tau$ (refs. 1-3). We assume $\tau$ to be same for all minibands and magnetic fields. In this approximation, $\sigma_{xx} \propto v^2 \tau$ and is determined by the group velocity of charge carriers, $v$ (ref. 26). Each BZ miniband effectively represents a different two-dimensional system with a different $k$-dependent velocity. If $T$ is larger than the cyclotron gaps, as in our case, the Fermi step becomes smeared over several minibands, which all contribute to $\sigma_{xx}$. In this regime, one can define $\langle v^2 \rangle$ averaged over an interval of $\pm T$ around $\varepsilon_F$. We calculate $\langle v^2 \rangle$ using a representative miniband spectrum for a graphene-on-hBN superlattice, which was computed using the model developed in ref. 30. The resulting conductivity is evaluated as[26]

$$\sigma_{xx} = \frac{4e^2}{h} \frac{\pi \varepsilon_F \tau}{h} \frac{\langle v^2 \rangle}{v_F^2},$$

where $v_F$ is the Fermi velocity and $h$ the Planck constant. The only fitting parameter is $\tau$, which we choose so that $\sigma_{xx}$ fits the experimental values for $\phi = \phi_0/2$ (Fig. 3). For other $p/q$, the calculated $\sigma_{xx}$ are shown by black dots. Furthermore, according to the classical magnetotransport theory[1-3], $\sigma_{xx}$ near zero $B_{eff}$ should vary as $\sigma_{xx}(B_{eff}) = \sigma_{xx}(0) - \alpha B_{eff}^2$ where $\alpha$ is a $p$ and $q$ -dependent coefficient. It can be evaluated[26] without extra fitting parameters (dotted parabolas in Fig. 3). One can see that the theory provides qualitative agreement for the amplitude and width of the experimental peaks. The derived values of τ yield $\sigma_{xx}(B = 0)$ ≈20 mS, again in qualitative agreement with experiment. It would be unreasonable to expect any better agreement because of the limited knowledge about the graphene-on-hBN superlattice potential[20,30] and the used τ-approximation. The observed exponential *T* dependence of BZ oscillations (fig. S4) can also be understood qualitatively as arising from scattering on acoustic phonons such that the scattering length ($\propto 1/T$) becomes shorter than the characteristic size, *aq*, of supercells responsible for the *q*-peak in conductivity[26].

To conclude, graphene superlattices exhibit a new quantum oscillatory phenomenon that can be understood as repetitive formation of distinct metallic systems, the Brown-Zak minibands. At simple fractions of $\phi_0$, charge carriers effectively experience zero magnetic field, which results in straight rather than curved (cyclotron) trajectories. Straighter trajectories lead to weaker Hall effect and higher conductivity. The background magnetoresistance is attributed to trajectories that involve transitions between different minibands and effectively become curved. The reported oscillations do not require the monochromaticity, which allows them to persist up to exceptionally high *T*, beyond the existence of Landau levels. The extrapolation of the observed *T* dependences (fig. S4b) suggests that the quantum oscillations may be observable even at 1,000 K. Further theory is required to understand details of temperature, field and concentration dependences of BZ oscillations, the origin of the electron-hole asymmetry of phonon scattering, behavior of higher-order fractions and the effect of inter-miniband scattering, which is responsible for the non-oscillating background.

## Supplementary Information

**S1 Device fabrication**

Our devices were made following procedures similar to those described previously[31]. First, an hBN/graphene/hBN stack was assembled using the dry-peel technique[32]. This involved mechanical cleavage of graphite and hBN on top of an oxidized silicon wafer, after which graphene monolayers and relatively thick (30-70 nm) hBN crystallites were identified using an optical microscope. The crystals were picked up from the substrate using a polymer membrane attached to the tip of a micromanipulator to assemble a three-layer stack such that graphene was encapsulated between two hBN crystals. Both graphene and hBN cleave preferentially along their main in-plane crystallographic axes, which often results in crystallites having relatively long and straight edges. Such edges were used to align graphene with respect to the bottom hBN using a rotational stage and by controlling the procedure under an optical microscope. The resulting accuracy of alignment was $\approx 1.5°$ (ref. 33). The top hBN crystal was misaligned intentionally to avoid possible contribution from a competing moiré potential[18,20]. The resulting heterostructure was placed on top of an oxidized silicon wafer (*n*-doped Si with $SiO_2$ thickness of either 90 or 290 nm) which served as a back gate. The next step involved electron beam lithography to make windows in a polymer mask, which defined contact regions to graphene. Reactive ion etching was employed to mill trenches in the heterostructure using these windows, which exposed the graphene monolayer. Metallic contacts (3 nm Cr/ 80 nm Au) were evaporated into the trenches, which was followed by lift-off of the polymer mask. This procedure prevented contamination of exposed graphene edges with polymer residues, resulting in high-quality contacts[31]. Finally, another round of lithography and ion etching was used to define a device in the multiterminal Hall bar geometry (fig. S1). The Hall bars had typical widths of 1–4 µm and lengths 10–20 µm.

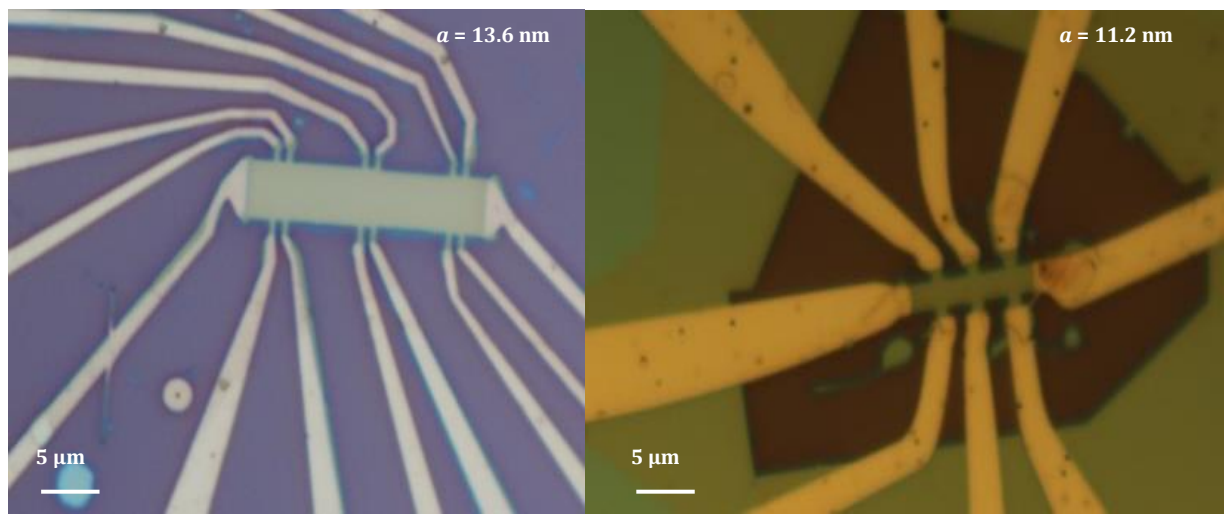

**Figure S1| Experimental devices**. Optical images of two of our devices. Their superlattices had periodicities *a* specified on the photos, as evaluated from the position $n_0$ of the second-generation NPs.

## S2 Further examples of Brown-Zak oscillations

BZ oscillations are found to be a universal feature of graphene superlattices. To show this, fig. S2 mirrors the presentation of Figs. 1a-b but uses a different superlattice device with a shorter $a$ (that is, larger $n_0$). Again, one can see that, in the reference device, quantum oscillations become rapidly suppressed and practically disappear already at 50 K. In contrast, the graphene superlattice exhibits oscillations that remain clearly visible at room $T$ in fields as low as 2.5 T. For this particular carrier concentration $n \approx 0.6n_0$, the amplitude of oscillations varies non-monotonically with increasing temperature. Below 3 T, the SdH oscillations are completely washed out at 50 K whereas the BZ oscillations emerge only at 100 K, seemingly growing with increasing $T$. We attribute this nonmonotonic dependence to beatings between the SdH and BZ oscillations. On the Landau fan diagrams this corresponds to fields and concentrations where minima in $\sigma_{xx}$ due to Landau quantization cross the conductivity maxima due to the formation of BZ minibands (see Fig. 2).

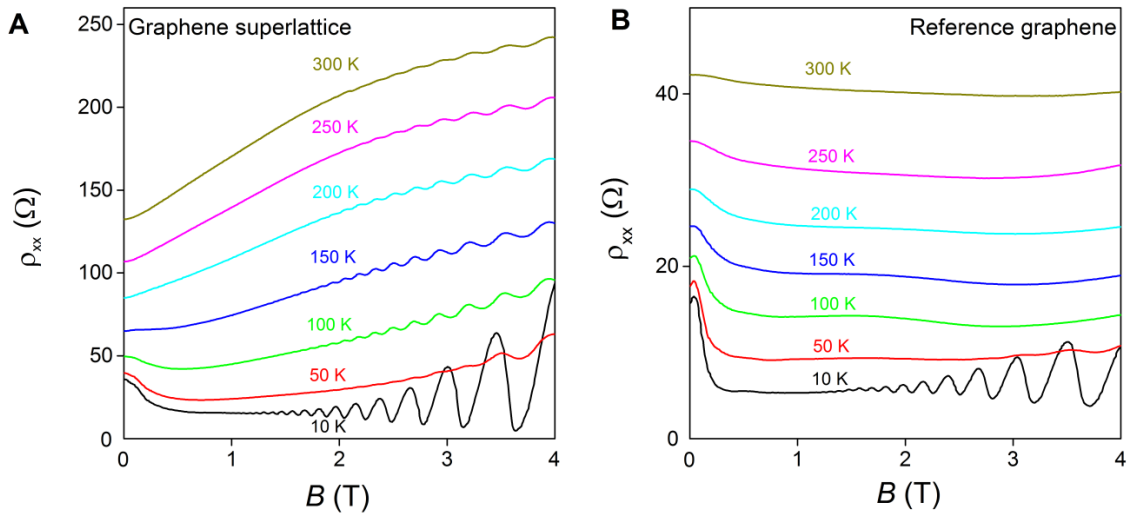

**Figure S2|** Emergence of Brown-Zak oscillations with increasing temperature. **a,** $\rho_{xx}(B)$ for a graphene superlattice with $a \approx 11.2$ nm ($n_0 \approx 3.5 \times 10^{12}$ cm$^{-2}$) at various $T$. **b,** Same for the reference graphene device. For both plots, $n \approx 2.2 \times 10^{12}$ cm$^{-2}$.

## S3 Frequency of Brown-Zak oscillations

BZ oscillations are periodic in $1/B$ (Figs. 2e-f & figs. S3-S4), so that at small amplitudes they can be described by $\cos(2\pi B_0/B)$. Their frequency $B_0$ was found to be independent of $n$ but varied from sample to sample (fig. S3b). This is attributed to different periodicities $a$ of the moiré pattern in different devices, which is caused by random, slightly different alignment between the crystallographic axes of the graphene and bottom hBN lattices[17-25]. For our devices, $a$ varied between $\approx$11.2 and 14.2 nm and could readily be evaluated from the carrier concentration $n_0$ at which second-generation NPs occurred. Indeed, the NPs occur if the lowest electron or highest hole minibands are fully occupied, which correspond to 4 charge carriers per superlattice unit cell with area $S$ (refs. 17-25, 30 & 34). This leads to the equation $n_0 = 4/S = 8/\sqrt{3}a^2$, which relates $n_0$ and $a$.

Examples of BZ oscillations with different frequencies are shown in fig. S3. One can see that the oscillations are fastest (largest $B_0$) for the device with the shortest period $a$ (largest $n_0$). The inset of fig. S3b summarizes the observed behavior $B_0(n_0)$ using data for all the studied devices. These are the same data set as in Fig. 2c of the main text but the specific devices from figs. S3a-b are now indicated by arrows. The experimental dependence is accurately described by the equation $B_0 = \phi_0/S$ or, equivalently, $B_0 = \phi_0 n_0/4$. This means that maxima of BZ oscillations in $\sigma_{xx}$ occur exactly at $B = \phi_0/qS$.

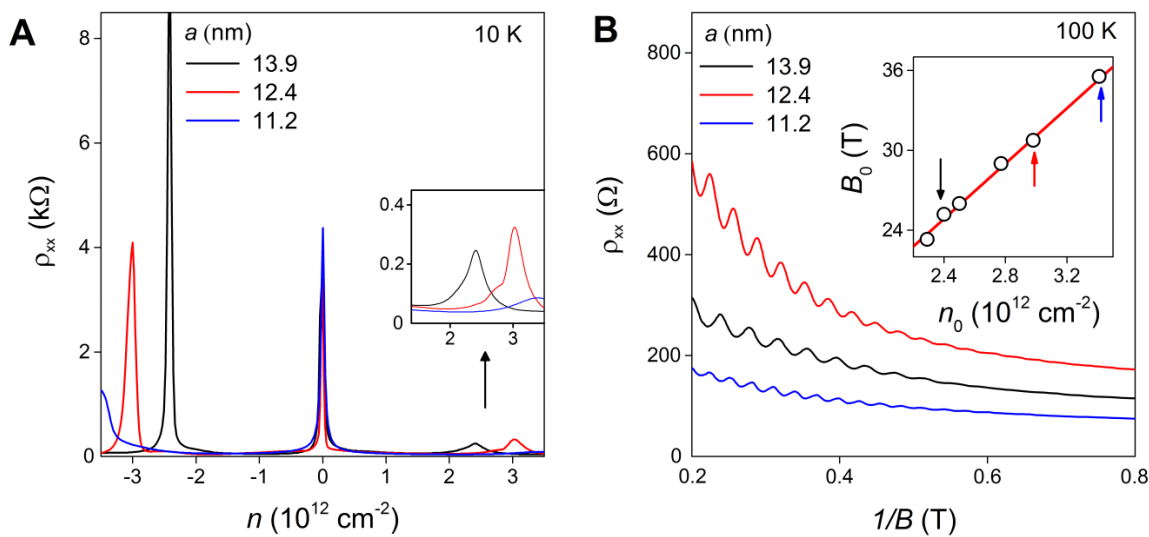

**Figure S3| Frequency of Brown-Zak oscillations for different graphene superlattices. a,** Examples of characterization of our superlattice devices. The second-generation NPs are found at $\pm n_0$ which can be translated into the values of $a$ and $S$ for the superlattices[17-25]. Inset: Behavior near the electron NPs is magnified. **b,** BZ oscillations for the three devices in (a) at approximately the same normalized density $n/n_0 \approx 0.75$. Inset: Frequency $B_0$ of the BZ oscillations for all our superlattice devices as a function of the position of their NPs (symbols). This includes the devices in (a,b) as indicated by the color-coded arrows. Red line: Expected dependence $B_0 = \phi_0 n_0/4$. $B_0$ corresponds to the last maximum in $\sigma_{xx}$ that should appear at one $\phi_0$ per superlattice unit cell.

**S4 Amplitude of BZ oscillations**

At low $T$ and high $B$, BZ and SdH oscillations coexist, which not only makes it difficult to find out a functional dependence for BZ oscillations but also leads to such abnormalities as the nonmonotonic $T$ dependence of the overall amplitude of quantum oscillations, as discussed for the case of fig. S2. Therefore, to examine behavior of BZ oscillations quantitatively, we have focused on high $T$ and relatively low $B$ where the contribution of SdH oscillations is small. In addition, SdH oscillations in graphene are suppressed with increasing $n$ because cyclotron gaps become smaller with increasing $\varepsilon_F$ (refs. 28 & 29). Accordingly, we studied functional dependences of BZ oscillations at $n$ above the second-generation NP. In this regime, BZ oscillations also become stronger (Fig. 2c, fig. S6) so that we could accurately measure their amplitude over a wide range of $T$. Examples of the observed $T$ and $B$ dependences are shown in fig. S4. The measured amplitude of the BZ oscillations, $\Delta\sigma_{xx}$, is plotted in fig. S4b as a function of $T$ for different $B$ up to 5 T. This covers a range of rational fluxes described by $q$ from 5 up to 37. One can see that $\Delta\sigma_{xx}$ changes exponentially over two orders of magnitude, suggesting the functional dependence $\Delta\sigma_{xx} \propto \exp(-T/T^*)$ where $T^*$ is a constant.

With regard to the field dependence, fig. S4b shows that the BZ oscillations decay slower with $T$ for higher $B$. For $T > 200$ K where the effect of cyclotron gaps is negligible, the BZ oscillations became small enough and practically sinusoidal. In this regime, we find that their amplitude $\Delta\sigma_{xx}$ can be described accurately as an exponential dependence on $1/B$ (fig. S4c). Therefore, the experiments suggest that the amplitude of BZ oscillations has the functional form $\ln(\Delta\sigma_{xx}) \propto -T/B$.

Qualitatively, this dependence can be understood as follows. BZ oscillations arise due to spatial quantization at the length scale $L \approx aq \equiv aB_0/B$, which involves $q$ unit cells in the makeup of the magnetic miniband arising for $B = \phi_0/qS$ (refs. 7, 8, 30 & 34). As long as Bloch wavefunctions of this miniband are not completely randomized by scattering, the miniband electronic structure is expected to affect transport properties of a graphene superlattice. Over the $T$ range of our experiments, electron collisions on acoustic phonons are known to be the dominant scattering mechanism. It is described by a mean free path $\ell \propto 1/T$. For small amplitude of oscillations, it is reasonable to assume that $\Delta\sigma_{xx}$ is an exponential function of the dimensionless parameter, $L/\ell$, which translates into $\ln(\Delta\sigma_{xx}) \propto -aTB_0/B \propto -T/B$, in agreement with the experiment. Further theoretical analysis and modeling are required to explain the observed $T$ and $B$ dependence quantitatively.

As discussed in the main text, we had to limit the $T$ range in our experiments in order to avoid irreversible damage of our devices. Nonetheless, let us note that, by extrapolating the dependences of fig. S4b to higher $T$ (for example, consider the 5T curve in this figure), one can find that BZ oscillations should in principle be observable up to 500 K ($\Delta\sigma_{xx}$ extrapolates to >0.01 mS). This consideration agrees well with Fig. 1c, where the oscillations are clearly seen at 373 K below 4 T. Moreover, in the latter case, $n < n_0$ and, therefore, BZ oscillations are weaker than those for $n$ above the second-generation NP (Fig. 2C). The observed $T/B$ functional form for the oscillation amplitude suggests that increasing $B$ to 10 T (by a factor of 2) should result in the same $\Delta\sigma_{xx} > 0.01$ mS at 1,000 K. However, electron-electron scattering increases with temperature as $T^2$ and, for graphene, the corresponding mean free path is expected to reach a 10-nm scale close 1,000 K, too. Therefore, we conservatively estimate that the widely available magnets with $B = 16$ T should allow the observation of BZ oscillations at approximately 1,000 K, assuming that no additional scattering mechanism is unexpectedly activated above 400 K.

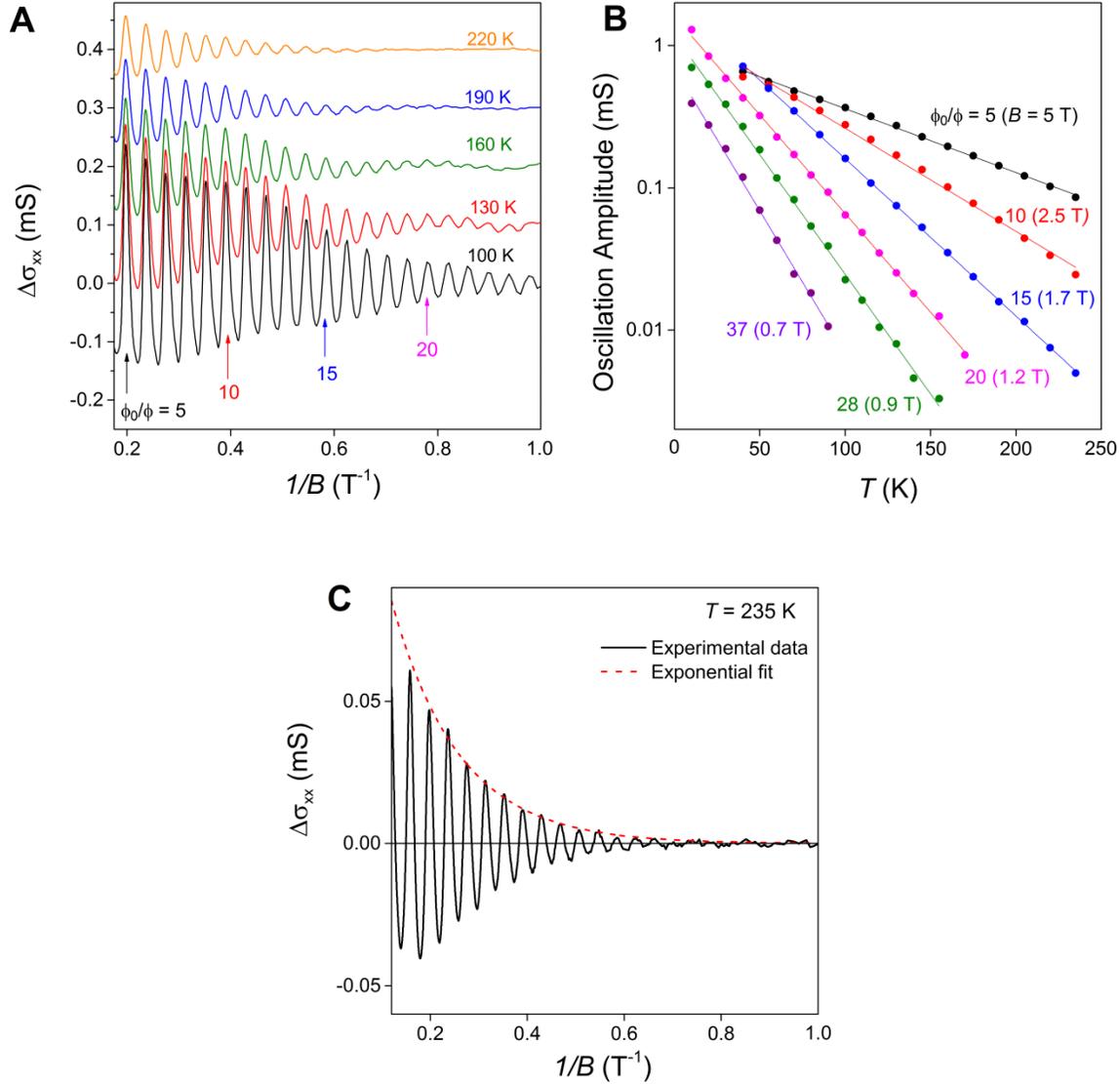

**Figure S4| Temperature and field dependence of Brown-Zak oscillations**. **a,** Examples of $\Delta\sigma_{xx}$ as a function of $1/B$ at different $T$ for our superlattice device with $a \approx 13.6$ nm ($n_0 \approx 2.5\times10^{12}$ cm$^{-2}$). The presented curves are for $n \approx 2.9\times10^{12}$ cm$^{-2}$. For clarity, the curves are shifted vertically by 0.1 mS. **b,** Detailed $T$ dependence of the oscillation amplitude. Some of the maxima are marked by their $q$ and color-coded in **a**. **c,** At high $T$, BZ oscillations decay exponentially as a function of $1/B$.

**S5 Density of states in graphene superlattices**

Graphene devices are known to exhibit strong dependence of their differential capacitance $C$ on gate voltage, which is due to both graphene's low DoS and the use of nanometer thin dielectric layers between graphene and the gate[20,35,36]. The latter minimizes a contribution from the classical (geometric) capacitance making it easier to measure the quantum capacitance due to the varying DoS. Capacitance measurements were previously employed to study the DoS in pristine graphene[35,36] and, more recently, in graphene superlattices at low $T$ (ref. 20). Using the same technique and experimental procedures as those described in ref. 20, we studied the DoS for several superlattice devices. Examples of the measured Landau fan capacitance diagrams $C(B,n)$ are shown in fig. S5. Cyclotron gaps appear in the diagrams as black stripes (minima in $C$). At low $T$, LLs are seen fanning out from the main and second-generation NPs. The latter cyclotron gaps are more pronounced for holes than electrons. Intersections of main and second-generation LLs result in third-generation NPs, near which the replica quantum Hall ferromagnetism was reported[20]. The low-$T$ behavior of the DoS in graphene superlattices (fig. S5a) agrees well with that reported previously (revealing the Hofstadter butterfly spectrum) and, therefore, we refer to ref. 20 for further explanations.

Taking into account that, in transport experiments, BZ oscillations become better resolved at elevated $T$, we have extended the capacitance measurements into the high-$T$ regime. As expected, cyclotron gaps in the DoS become smeared with increasing $T$. Only the largest gaps for the main spectrum plus the DoS minima originating from the second-generation NP for hole doping could be observed at 100 K (fig. S5b). Importantly, no sign of horizontal streaks can be discerned in fig. S5, not only at high but also at low $T$. This is in stark contrast to transport measurements where such streaks that correspond to BZ oscillations are always present on Landau fan diagrams (Fig. 2, fig. S7a).

In more detail, fig. S5c shows $C(B)$ curves at several $T$ for fixed $n$ near the second-generation NP for electrons, where BZ oscillations are strongest in transport experiments. No oscillatory behavior is noticeable in the capacitance measurements above 100 K. At 200 K, there is no sign left of the LL minimum even at 15 T. For comparison, fig. S5d plots $\rho_{xx}$ measured at the same $T$: BZ oscillations appear already at $B < 3$ T. To summarize, no signatures of $n$-independent oscillations were found in the DoS at any $T$ and $B$, which strongly suggests that the high-$T$ oscillations observed in transport characteristics are not due to any extra gaps in the electronic spectra of graphene superlattices.

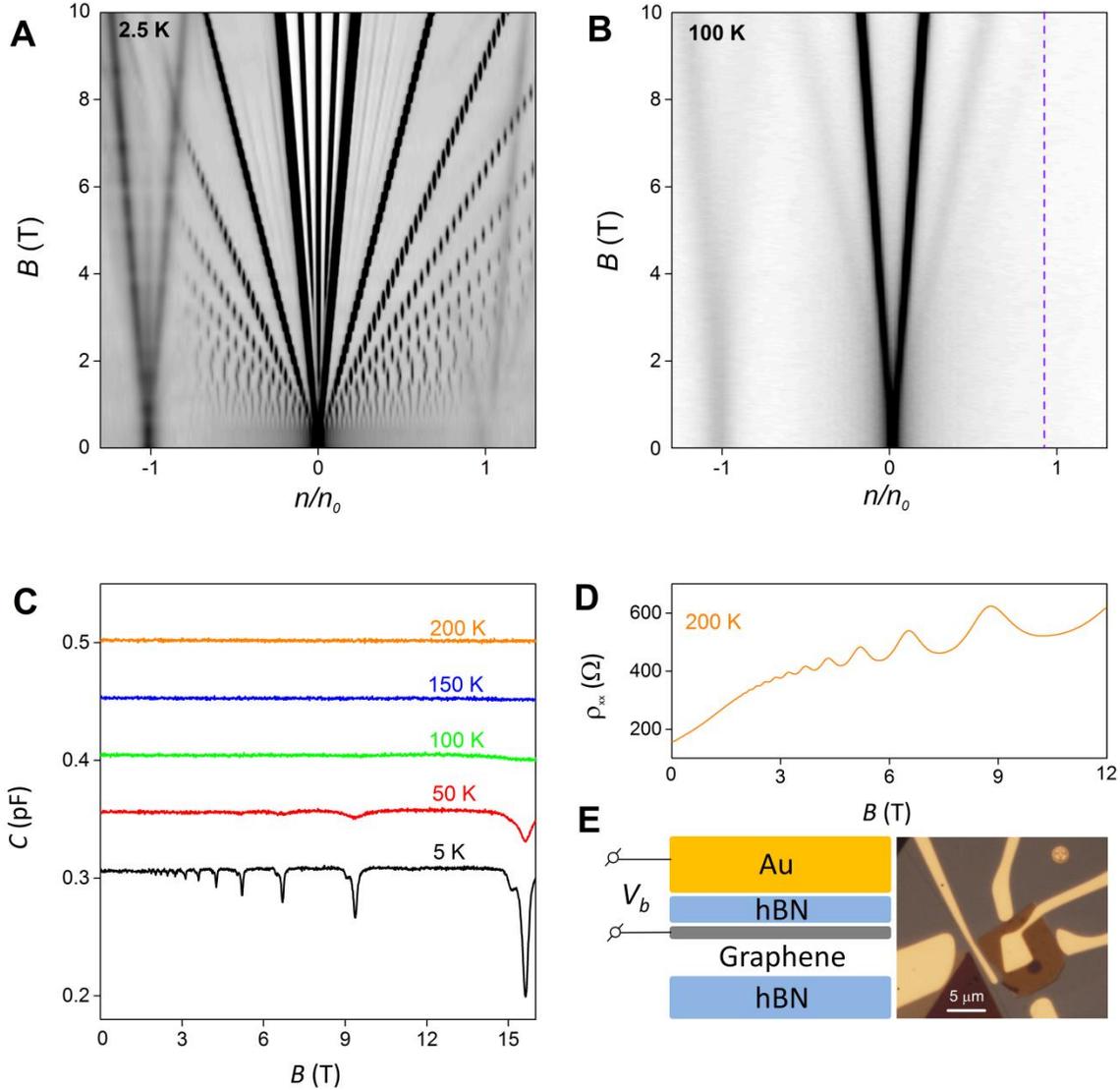

**Figure S5| Differential capacitance for graphene superlattices. a,** Low-$T$ Landau fan diagram $C(n,B)$ for a capacitor with $a \approx 13.5$ nm. Grey linear scale: black 0.266 pF; white 0.293 pF. **b,** Same as in (a) but at 100 K. Scale: black 0.328 pF; white 0.357 pF. **c,** Examples of $C(B)$ for the fixed $n \approx 2.3\times10^{12}$ cm$^{-2}$ ($n/n_0 \approx 0.9$) which corresponds to the purple dashed line in (b). The curves are shifted vertically for clarity. **d,** $\rho_{xx}(B)$ for a Hall bar device with $a \approx 13.9$ nm and measured at the same $n$ and $T$ as the orange curve in (c). **e,** Schematic and optical image of one of our capacitance devices. For details, see ref. 20.

## S6 Higher-order BZ oscillations

At electron concentrations beyond the second-generation NP, additional features were found in the transport properties of graphene superlattices at elevated $T$. Extra maxima could be discerned in Fig. 2e (arrows) but they are even better resolved in fig. S6a where $n$ was increased to $\approx 2n_0$, close to the highest concentration accessible for our devices. In addition, fig. S6b plots both $\sigma_{xx}$ and $\sigma_{xy}$ in this regime at a slightly lower $T$ of 100 K to enhance the extra features. The maxima in $\sigma_{xx}$ align with smeared step-like features in $\sigma_{xy}$. Their positions correspond to $B = p\phi_0/qS$ where $p$ = 2 and 3, as indicated by the dotted curves and blue arrows, respectively. We attribute these transport anomalies to the formation of Brown-Zak minibands for the case where $p$ flux quanta penetrate through $q$ unit cells[7,8,30,34]. The features are weaker than those for unit fractions $\phi_0/q$ in the same range of $B$ because $p$ times larger areas are involved in their formation[7,8]. This is in agreement with the above discussion of the amplitude of BZ oscillations, which decays exponentially with the involved length scale $L \approx apq$.

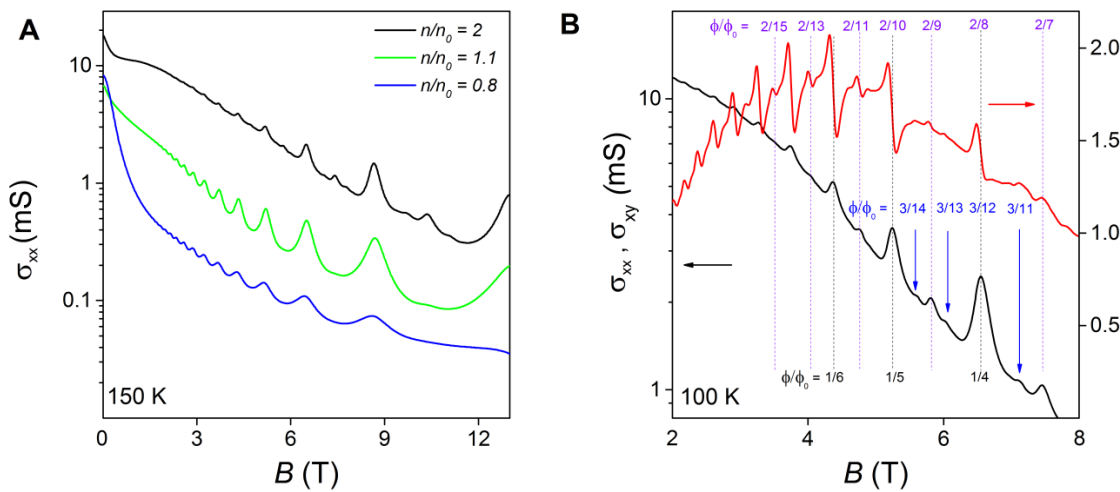

**Figure S6| Fractal BZ oscillations**. **a,** High-$T$ conductivity in graphene superlattices with increasing $n$ beyond the second-generation NP for electrons. The same device as in Fig. 2 ($a \approx 13.9$ nm). New features are seen to appear on the curve with $n = 2n_0$. The logarithmic scale is used to amplify the additional maxima in $\sigma_{xx}$ against the background. **b,** Comparison of the extra features in $\sigma_{xx}$ with those in $\sigma_{xy}$. The electron density is $2n_0$.

**S7 BZ oscillations for hole doping**

It is well established that the DoS in graphene-on-hBN superlattices is modified stronger for the valence than conduction band[17,25,34]. The resulting electron-hole asymmetry is also clearly seen in our present devices where cyclotron gaps in the DoS (figs. S5a-b), the second-generation NPs (fig. S3a) and LLs in $\sigma_{xx}$ at low $T$ (Fig. 2a) are most pronounced for hole doping. In contrast, BZ oscillations are more visible for electrons rather than holes (Fig. 2b) and, accordingly, the main text focused on the results obtained for electron doping (positive $n$). To emphasize the generality of this 'reversed' electron-hole asymmetry, figs. S7a-b show data similar to those of Fig. 2 but for another superlattice device ($a \approx$ 13.6nm).

From the experimental point of view, the origin of the unexpected asymmetry becomes clear if we look closer at the $T$-dependence of scattering in graphene-on-hBN superlattices. To this end, fig. S7c plots $\rho_{xx}$ as a function of electron and hole doping at different $T$. The resistivity grows faster with increasing $T$ for the valence band compared to the conduction band. The behavior is quantified in fig. S7d using the fixed concentrations $n/n_0 \approx \pm 0.7$ for electrons and holes. One can see that electron-phonon scattering evolves linearly with *T* and is approximately 4 times stronger for holes than electrons. It is also clear from fig. S7d, that superlattice effects strongly enhance phonon scattering with respect to pristine graphene. Indeed, the reference devices (encapsulated but non-aligned graphene) exhibit one-to-two orders of magnitude weaker phonon scattering. Because the amplitude of BZ oscillations depends exponentially on the mean free path $\ell$ (fig. S4), it is hardly surprising that, at elevated *T*, the oscillations are more strongly suppressed for hole doping. The asymmetry of electron-phonon scattering is apparently caused by the moiré pattern and has not been noted previously[17-25]. The asymmetry's origin is likely to be specific to graphene-on-hBN superlattices and remains to be understood.

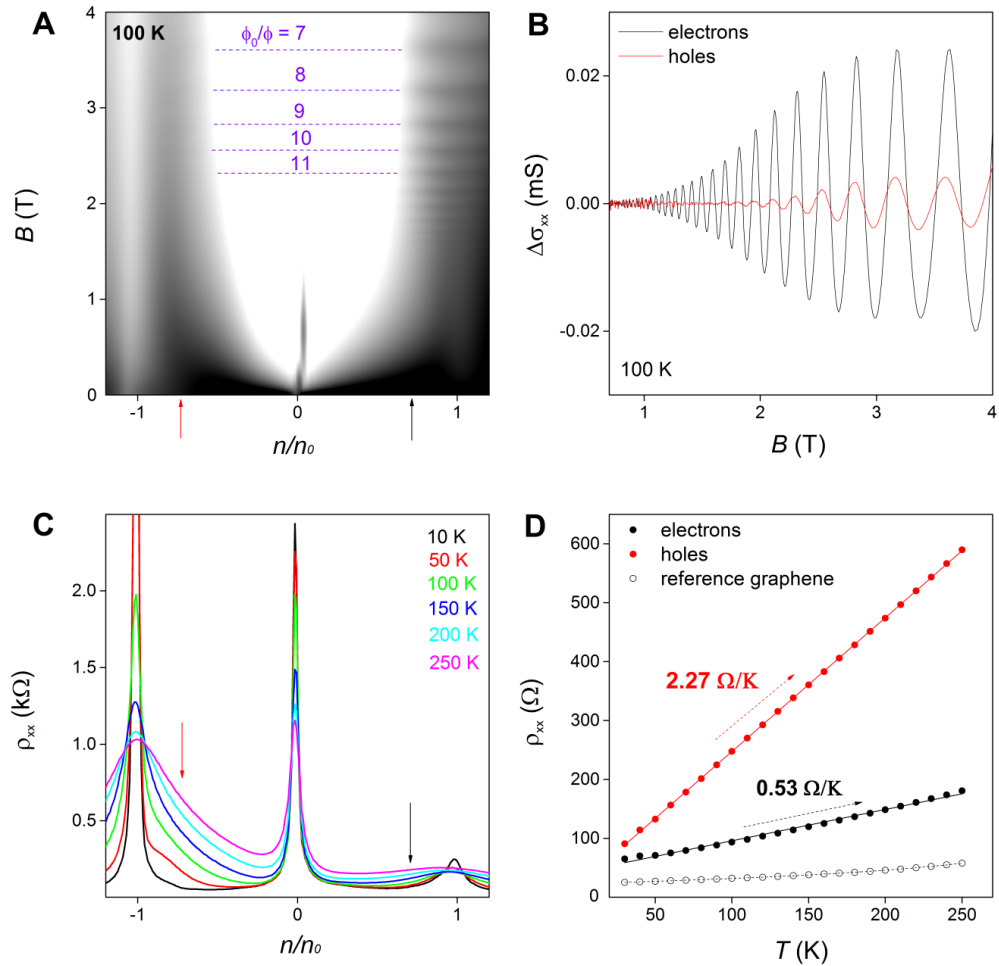

**Figure S7| Electron-hole asymmetry of Brown-Zak oscillations**. **a,** $\sigma_{xx}$ as a function of *n* and *B*, similar to Fig. 2B but for a superlattice device with $a \approx 13.6$ nm. The grey scale is logarithmic: white 0.09mS; black 5mS. **b,** After subtracting a smooth background, $\Delta\sigma_{xx}(B)$ is plotted for the fixed $n \approx \pm 1.7 \times 10^{12}$ cm$^{-2}$ [$n/n_0 \approx \pm 0.7$]. These electron and hole concentrations are marked by the black and red arrows in (a), respectively. **c,** $\rho_{xx}$ for the same device as a function of *n* at various *T* in zero *B* (the curves are color coded). The resistivity increases with *T* faster for holes than electrons and much faster than in reference devices. **d,** Detailed *T*-dependence for electrons and holes at the same *n* as in (b). Dots – experiment;

lines – best linear fits. Values for the linear slopes are shown next to the curves. For comparison, $T$-dependence found in the reference device is plotted for the same concentration (open circles). In non-aligned devices, electrons and holes exhibit similar $T$-dependences[28, 29].

**S8 Brown-Zak magnetic minibands**

Let us consider graphene's spectrum modified by the presence of a moiré potential induced by the hBN substrate. We use a hexagonal Bravais lattice $n_1\vec{a}_1 + n_2\vec{a}_2$ with the superlattice period $a = a_1 = a_2$. In magnetic fields, $B = B_{p/q} = p\phi_0/qS$, magnetic minibands are formed[7,8]. To describe these minibands, we employ the phenomenological model developed in[30,34] and based on the Hamiltonian

$$\hat{H} = v_F \vec{p}\cdot\vec{\sigma} + u_0^+ f_+ + \xi\sigma_3 u_3^+ f_- + \frac{\xi}{b} u_1^+ \vec{\sigma}\cdot[\vec{\ell}_z \times \nabla f_-],$$

$$f_\pm = \sum_{m=1\ldots 6}(\pm 1)^{m+\frac{1}{2}} e^{i\vec{b}_m\cdot\vec{r}}, \quad \text{(S1)}$$

where $\sigma_i$ are the Pauli matrices acting on the sublattice Bloch states $(\phi_{AK}, \phi_{BK})^T$ in the K valley ($\xi = 1$) and $(\phi_{BK'}, -\phi_{AK'})^T$ in the K' valley ($\xi = -1$). $f_\pm$ are six Bragg vectors $\vec{b}_m$ ($b_{1,2,3,4,5,6} = b = \frac{4\pi}{\sqrt{3}a}$) of the superlattice. The effect of magnetic field is incorporated in (S1) as $\vec{p} = -i\hbar\nabla + e\vec{A}$ where the vector potential $\vec{A} = \frac{Bx_1}{a\sqrt{3}}(2\vec{a}_2 - \vec{a}_1)$ is written in the hexagonal (non-orthogonal) coordinate system $(x_1, x_2)$ such that $\vec{r} = x_1\vec{a}_1 + x_2\vec{a}_2$, which was adapted for the case of hexagonal superlattices[34].

The miniband spectrum plotted in fig. S8 was obtained using the computational procedure developed in ref. 34. In our modelling, we chose $u_0^+ = 21.7$ meV, $u_1^+ = -30.6$ meV and $u_3^+ = -22.2$ meV, which are to be compared with $v_g b_m = 357$ meV, the case of perfectly matching graphene and hBN lattices. The resulting spectrum displays the same features as discussed earlier[34,37-42] using various models for graphene-hBN coupling:

i. At $\phi \ll \phi_0$, minibands converge into LLs separated by the cyclotron gaps $\sim\hbar\omega_c$ which are large for massless Dirac fermions in graphene.
ii. Individual minibands are systematically wider at unit fractions $\phi = \phi_0/q$ compared to nearby rational flux values. Minibands are $q$-fold degenerate and dispersed over a Brillouin zone with the area, $S_{BmZ} = \frac{8\pi^2}{3\sqrt{3}}(qa)^{-2}$.
iii. If intervals $\frac{1}{q+1}\phi_0 < \phi < \frac{1}{q}\phi_0$ for different $q$ are compared, the sparsity of the spectrum increases upon increasing $q$. Minibands for unit fractions $1/q$ are widest, followed by the sequence $2/q$ that exhibits somewhat narrower bands and, then, by the $3/q$ sequence.
iv. Away from the unit fractions, minibands bunch so that they can be described as LLs for third-generation Dirac fermions[20,34]. These LLs are effectively the response of the BZ minibands to the effective magnetic field $B_{eff} = B - B_{1/q}$, with the gaps set by the corresponding effective cyclotron energies $\sim\hbar\omega_c^{eff}(B_{eff})$.

The spectrum in fig. S8 is generic[30,34], and we use it below to explain only the qualitative features expected for electron transport in graphene superlattices.

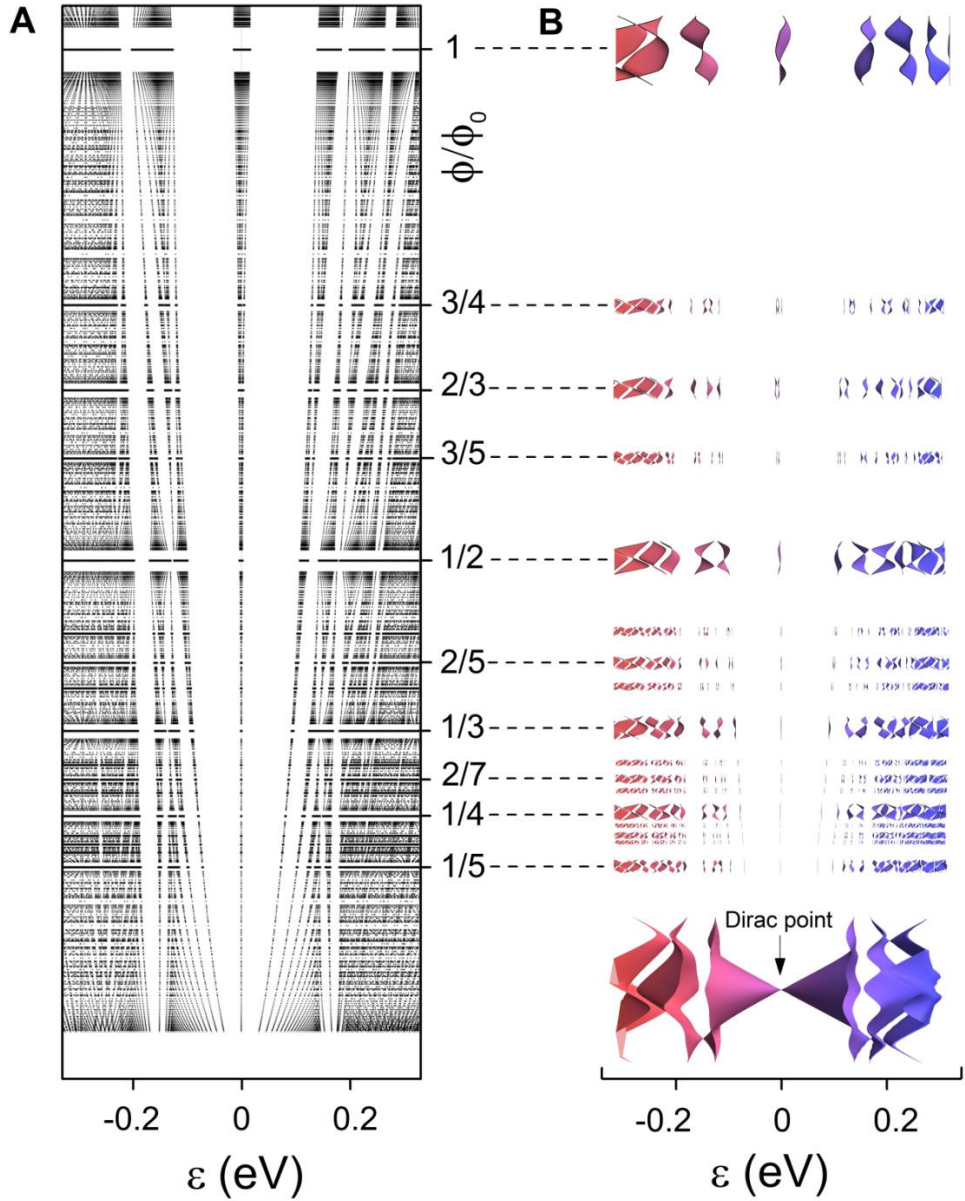

**Figure S8| Spectra of graphene superlattices in magnetic field. a,** Bands and gaps in the Hofstadter spectrum. Black dots: States calculated for numerous simple fractions *p/q*. The horizontal white stripes correspond to missing data because of computational limitations [for details, see refs. 30 & 34)]. **b,** Dispersions $\varepsilon(\vec{k})$ in magnetic minibands that appear exactly at $\phi/\phi_0 = p/q$ for some $q = 1$ to 5 and $p = 1$ to 3. The electronic spectra are aligned vertically against the corresponding values of $\phi/\phi_0$ in (a) but show the dispersion against $k_x$ and $k_y$ within the first Brillouin minizones (similar to the case of the inset in Fig. 3). The lowest inset to the right is the superlattice's modified Dirac spectrum in zero B (ref. 30).

## S9 Magnetotransport in Brown-Zak minibands

At low temperatures, $kT \ll \hbar\omega_c, \hbar\omega_c^{eff}$, the hierarchy of gaps in the Hofstadter spectrum manifests itself in a sequence of incompressible quantum Hall states, which were studied earlier using magnetotransport and magnetocapacitance measurements[17-22]. High temperatures, $kT \gg \hbar\omega_c$, smear the Fermi step over several minibands, and this makes measurements insensitive to the presence of even large spectral gaps, leaving aside extra gaps due to the superlattice potential. In the high-$T$ regime, magnetotransport reflects the hierarchy of the width of magnetic minibands. Indeed, elastic diffusion of electrons in wider minibands should generate larger $\sigma_{xx}$ than in narrower minibands, resulting in magneto-oscillations periodic in $1/B$ where maxima in $\sigma_{xx}$ occur at $B = B_{\frac{1}{q}}$, the fields with widest minibands. Additional weaker maxima appear for $p > 1$, where minibands exhibit weaker broadening. The resulting oscillations can sustain higher $T$, compared to SdH oscillations, and be less sensitive to charge carrier inhomogeneity because they rely not on cyclotron gaps but, importantly, on the stability of the magnetic miniband structure and Bloch wavefunctions.

To describe the high-$T$ oscillations, we use the τ-approximation with a single τ and analyse the Boltzmann equation

$$\frac{\partial f}{\partial t} + e(\vec{E} + B\vec{l}_z \times \vec{v}) \cdot \nabla_p f + \vec{v} \cdot \nabla f = -\frac{1}{\tau}[f - f_F] \tag{S2}$$

where the occupancy for the plane-wave states $\vec{p}$ across Brillouin minizones with each '$n$-th' BZ miniband for a particular fraction $p/q$ is described by the distribution function $f(\vec{p}, n)$. In the linear response regime such that $f - f_F \propto E$ (where $f_F$ is the Fermi function at the base $T$), analysis can be performed using the Taylor expansion in powers of the effective magnetic field $B_{eff} = B - B_{\frac{1}{q}}$. Therefore, we write $f - f_F = f_1^{(0)} + B_{eff} f_1^{(1)} + B_{eff}^2 f_1^{(2)}$. By solving eq. (S2) using iterations in powers of $B_{eff}$, we find that

$$f - f_F = -\tau e\vec{E} \cdot \vec{v}\partial_\epsilon f_F + e^2 B\tau \left(\vec{l}_z \times \vec{v}\right) \cdot \nabla_p \left(\tau e\vec{E} \cdot \vec{v}\partial_\epsilon f_F\right)$$

$$-e^3 B^2 \tau(\hat{l}_z \times \vec{v}) \cdot \nabla_p(\tau(\hat{l}_z \times \vec{v}) \cdot \nabla_p(\tau e\vec{E} \cdot \vec{v}\partial_\epsilon f_F)).$$

Using the relation between dissipative conductivity $\sigma \equiv \sigma_{xx} = \sigma_{yy}$ (as prescribed by the hexagonal symmetry) and the Joule heating, we obtain

$$E^2 \sigma = \vec{E} \cdot \vec{j} = 4e \sum_n \int_{BmZ} \vec{E} \cdot \vec{v} f(\vec{p}, n) \frac{d^2 p}{(2\pi\hbar)^2}, \tag{S3}$$

and express it in terms of the band structure parameters computed for a group of $N$ minibands in Brown-Zak spectrum at each $\phi = \frac{p}{q}\phi_0$

$$\langle F \rangle = \frac{1}{NS_{BmZ}} \sum_n \int_{BmZ} F(\vec{p}, n) \frac{d^2 p}{(2\pi\hbar)^2}. \tag{S4}$$

We then use average values of the relevant band parameters that emerge from the iterative solution of eq. (S2) and are evaluated for the numerically computed miniband spectra for the model in eq. (S1). Eq. (S4) can be further simplified using the relation

$$\sum_n \int_{\text{BmZ}} F(\vec{p}, n) \frac{d^2p}{(2\pi\hbar)^2} \to \langle F \rangle \int \gamma(\varepsilon) \partial_\varepsilon f_F d\varepsilon = \langle F \rangle \gamma(\varepsilon_F),$$

which is based on the fact that, at the energy scale extended over several minibands, the DoS for the 'smeared' spectrum can be approximated by the DoS in the unperturbed graphene, $\gamma(\varepsilon_F) = \frac{2\varepsilon_F}{\pi\hbar^2 v_F^2}$. This leads to

$$\sigma = \frac{2e^2}{h} \frac{\varepsilon_F \tau}{\hbar} \left[ \frac{\langle v^2 \rangle}{v_F^2} + e^2 B^2 \tau^2 \frac{\langle \sum_{i=x,y} ([\vec{v} \times \nabla_p]_z v_i)^2 \rangle}{v_F^2} \right] \quad (S5)$$

which takes into account all valley and spin states. We evaluate mean values, $\langle v^2 \rangle$ and $\langle \sum_{i=x,y} ([\vec{v} \times \nabla_p]_z v_i)^2 \rangle$, by averaging the computed values over several minibands as illustrated in the inset of Fig. 3 and in fig. S8.

It is interesting to note that BZ oscillations can also be expected in diamagnetic response of superlattices, not only in their electron transport. Indeed, each BZ miniband effectively represents a distinct metallic system and, therefore, should exhibit specific diamagnetism. By changing magnetic field, one can sample these different states and is expected to observe a varying diamagnetic response with the periodicity $B = p\phi_0/qS$. This would be a de Haas – van Alphen -like effect but without Landau quantization. Such an analogue of BZ oscillations in magnetization seems to have been observed in recent tight-binding calculations[43], reflecting the recurring formation of different Bloch states.

**S10 Why have Brown-Zak oscillations remained unnoticed until now?**

There have been a number of experimental reports studying magnetotransport properties in aligned graphene-on-hBN devices[17-22,25]. These included measurements at elevated $T$. We believe that the reason why these high-$T$ oscillations have not been noticed earlier is partly due to the way in which Landau fan diagrams and $B$ dependences are usually measured for graphene devices. This involves sweeping gate voltage whereas other variables such as $B$ and $T$ are fixed. From the experimental point of view, this approach is most convenient. However, in such measurements, it is also easy to miss even very strong BZ oscillations, as illustrated in fig. S9. At low $T$, $\rho_{xx}$ for both superlattice and reference devices in fig. S9 exhibits multiple minima (SdH oscillations). In contrast, only peaks at the NPs survive at high $T$. Otherwise, $\rho_{xx}$ curves are featureless for both devices, with no sign of BZ oscillations even though they are quite pronounced if $B$, rather than $n$, is swept at a given $T$ (cf. fig. S5d). This is because BZ minibands appear and disappear as a function of flux per superlattice unit cell and, unlike SdH oscillations, do not vary with $n$.

We expect that BZ oscillations are not unique to graphene-on-hBN and can be observed for other moiré superlattices. A particularly promising candidate is twisted bilayer graphene that was reported to exhibit clear superlattice effects (see, e.g., refs. 44-45).

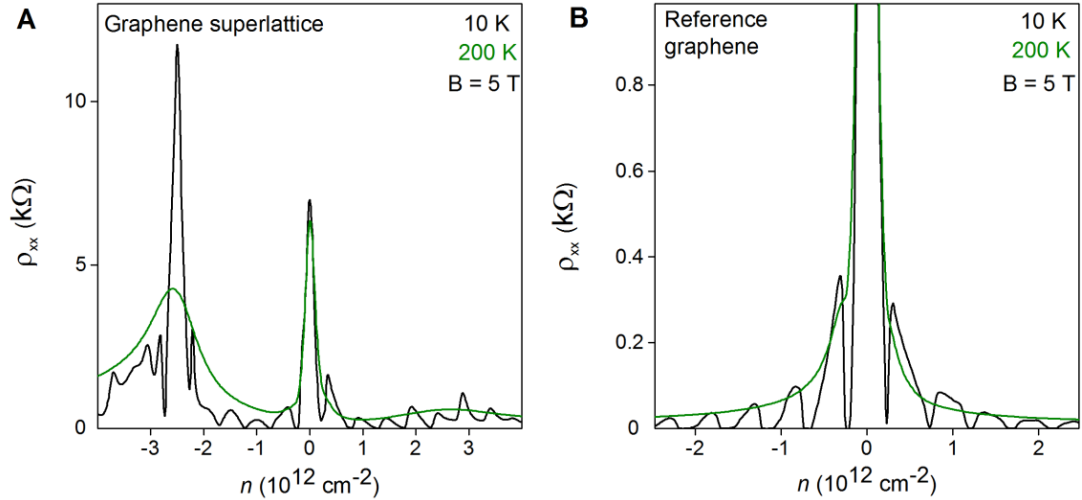

**Figure S9| Standard approach to measuring resistivity for graphene devices. (a** and **b)** Superlattice and reference devices, respectively. Landau quantization is clearly visible at 10 K but the curves become featureless at 200 K, except for the peaks at the NPs. BZ oscillations are quite strong in the superlattice device at 200 K for this range of *B* but do not show up at all in the measurements in (a) using charge-density sweeps.

**S11 Magnetotransport in low fields**

Fig. 2d shows that $\sigma_{xx}$ and $\sigma_{xy}$ exhibited rapid changes near zero *B*. These low-*B* features are irrelevant for the scope of the current report that focuses on BZ oscillations. Nonetheless, for completeness we show these features in more detail in fig. S10. It plots the experimentally measured $\rho_{xx}$ and $\rho_{xy}$ over the entire range of *B* in Fig. 2d and magnifies the behavior around zero *B*. The sharp dip in $\rho_{xx}$ and sign-changing $\rho_{xy}$ can be attributed to a complex electronic band structure of graphene superlattices at the energies above the second-generation NP (see the lowest panel in fig. S8b). The particular behavior (shown for $n/n_0 \approx 1.5$) was found to change with changing *n* only by a fraction of $n_0$. At high doping, several Brillouin minizones with opposite charge carriers are likely to contribute to the transport characteristics[20,30,34]. In addition, minibands can become depopulated with increasing *B*. Although the features in the inset of fig. S10 are relatively small, when translated into $\sigma_{xy}$, they lead to the notable nonmonotonic behavior at low *B* in Fig. 2d. This regime lies beyond the experimental range in which BZ oscillations are observable and requires separate investigation.

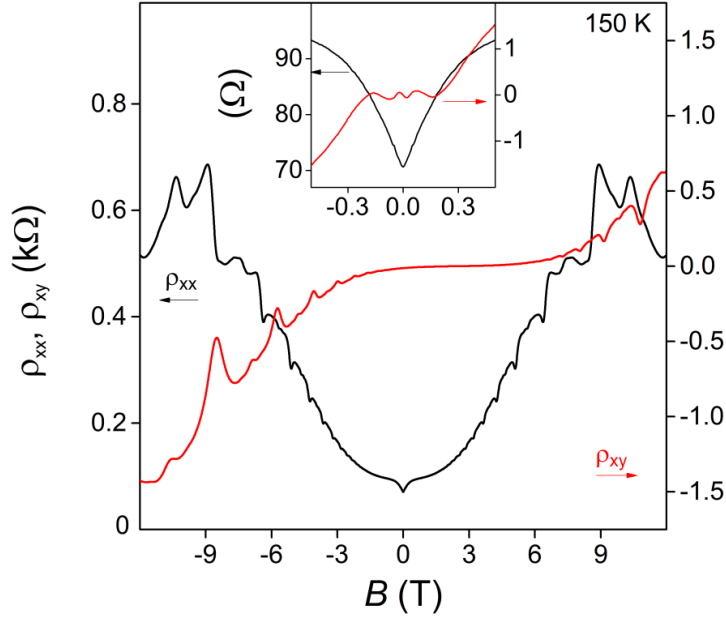

**Figure S10| Low-field transport at high doping**. $\rho_{xx}$ and $\rho_{xy}$ for the superlattice device in Fig. 2 of the main text. Both characteristics exhibit anomalous behavior in low $B$ such that the Hall effect changes its sign twice, and $\rho_{xx}$ shows a sharp dip. This is attributed to a complex miniband structure at energies beyond the second-generation NP. Inset: Low-$B$ region is magnified.